\documentstyle[aps,prl,psfig,multicol]{revtex}

\begin{document}

\title{First and second order clustering transitions \\
for a system with infinite-range attractive interaction}

\author{M. Antoni$^{(\$)}$, S. Ruffo$^{(\#,\,\&)}$ 
\footnote{INFM and INFN - Firenze}, 
and A. Torcini$^{(\#,\,\$)}$}

\address{
$^{\$}$ UMR-CNRS 6171 - 
Universit\'e d'Aix-Marseille III - 
Av. Esc. Normandie-Niemen, 13397 Marseille Cedex 20, France. \\
$^{\#}$ Dipartimento di Energetica ``Sergio Stecco'', Universit\'a
di Firenze, Via S. Marta 3, 50139 Firenze, Italy. \\
$^{\&}$ ENS-Lyon, Laboratoire de Physique, 46 All\'ee d'Italie,
Lyon Cedex, France}
\date{\today}
\maketitle

\begin{abstract}
We consider a Hamiltonian system made of $N$ classical particles 
moving in two
dimensions, coupled via an {\it infinite-range interaction} gauged by a
parameter $A$. This system shows a low energy phase 
with most of the particles trapped in a unique cluster.
At higher energy it exhibits a transition towards
a homogenous phase. For sufficiently strong coupling $A$ an 
intermediate phase characterized by two clusters appears. 
Depending on the value of 
$A$ the observed transitions can be either second or first 
order in the canonical ensemble. In the latter case microcanonical 
results differ dramatically from canonical ones. 
However, a canonical analysis, extended to metastable and unstable 
states, is  able to describe the microcanonical
equilibrium phase. In particular, a microcanonical negative specific heat 
regime is observed in the proximity of the transition whenever
it is canonically discontinuous. In this regime, {\it microcanonically 
stable} 
states are shown to correspond to {\it saddles} of the Helmholtz free 
energy, 
located inside the spinodal region.  
\end{abstract}

\vskip 0.5 truecm


\begin{multicols}{2}

It is well known that thermodynamic quantities derived 
within different statistical ensembles should coincide
in the thermodynamic limit. However, 
this statement is valid only if the interaction 
among the particles satisfies two conditions : 
{\it (i)} the pairwise interaction potential is integrable (
i.e it decays faster than $1/r^d$, where $d$ is the space dimension); {\it 
(ii)} the potential energy per particle is 
bounded from below \cite{kubo}. Whenever one of these conditions is 
violated,
ensemble inequivalence and thermodynamical instabilities can occur.
An extreme situation is represented by the gravitational 
potential for which neither condition is satisfied.
Indeed, for gravitating systems the usual laws of
equilibrium thermodynamics are expected not to hold : 
one of the most striking anomalies is related 
to the negative values taken by the specific 
heat~\cite{pad,lynden,chavanis}. 
Discrepancies between results obtained in the
microcanonical and
canonical ensembles, with an associated negative specific
heat regime, have been observed in the thermodynamic limit
for several systems with attractive potentials violating 
either condition {\it (i)} \cite{anttor,barre,ispo}
or condition {\it (ii)} only \cite{thir,compagner}.
Similar anomalies are also present for systems
with a finite number of particles, e.g. for nuclear 
multifragmentation~\cite{chomaz},
as well as for atomic clusters~\cite{labastie}.
In all these cases, preliminary results suggest that 
ensemble inequivalence can be observed in proximity of a 
canonically first order phase transition~\cite{barre,gross_book,leyvraz}. 

In this article, we aim at better clarifying the origin of
such inequivalence by considering a generalization of
a previously studied $N$-body classical Hamiltonian system
with infinite-range attractive interactions 
\cite{anttor}. The novelty of the present model consists in
the presence of a tunable coupling $A$ that allows to change the 
nature and the order of the transitions. These are investigated
analytically within the canonical ensemble and numerically 
via microcanonical molecular dynamics simulations.
This system shows three different phases :
a clustered phase ($CP_1$) occuring at low
internal energy $U$ (temperature $T$), with most of 
the particles trapped in a unique cluster,
a second clustered phase ($CP_2$), 
exhibiting two clusters, at
intermediate energy (temperature) 
and sufficiently strong coupling $A$,
and a homogenous phase ($HP$), with particles 
uniformly distributed, at high values of $U$ ($T$).
The canonical equilibrium solutions are computed from the lowest lying 
extrema of the
Helmholtz free energy and reveal that the system undergoes either first
or second order transitions depending on the value of the
coupling constant \( A \).  In particular, we have focused our attention 
on first order transitions separating the ordered phase $CP_1$ and 
the homogeneous one $HP$ and between the two clustered phases. 
In both cases canonical and microcanonical equilibrium 
predictions differ
dramatically near the transition, revealing a 
negative specific heat regime within the microcanonical ensemble.
No discrepancy between microcanonical and canonical
results appear at continuous phase transitions,
at least for what concerns the temperature-energy
equilibrium relation~\cite{tsa}.
 
Our results demonstrate that, irrespectively
of the nature of the phases involved in the transition,
the microcanonical negative specific heat regime can be 
well reproduced even within the canonical ensemble.
Once not only the absolute minima of the
Helmholtz free energy are taken in account, but also
the relative extrema corresponding to canonically
metastable and unstable states.                                 

The model we consider is a classical N-body Hamiltonian system 
defined on a two-dimensional periodic cell. The interparticle potential is 
infinite ranged and all the particles are identical with unitary mass. 
The Hamiltonian of the model is \( H_A 
= K + V_A
\) where $ K= \sum _{i=1}^{N}\left( 
\frac{p_{x,i}^{2}+p_{y,i}^{2}}{2}\right)$ is the kinetic energy 
and the potential energy reads as

\begin{eqnarray}
V_A&=&\frac{1}{2N}\sum^{N}_{i,j=1}\left[ 2+A-\cos 
\left( x_{i}-x_{j}\right) -\cos \left( y_{i}-y_{j}\right) \right. 
\nonumber \\
&&\left. -A\cos (x_{i}-x_{j})\cos (y_{i}-y_{j})\right] 
\quad ,
\label{eqHA}
\end{eqnarray}
with \( (x_{i},y_{i})\in ]-\pi :  \pi ] \times ]-\pi : \pi ] \) representing 
the coordinates of the $i$-particle and \( (p_{x,i},p_{y,i}) \) the 
conjugated momenta. 
For \( A=0 \), the two spatial directions $x$ and $y$ are uncoupled 
and the Hamiltonian reduces to the sum of two 
independent one-dimensional mean-field models~\cite{antruf}. 
In this case, as shown in \cite{antruf},
a second order phase transition appears, both in the microcanonical
and in the canonical ensemble, connecting a single clustered phase 
(\( CP_1 \)), at sufficiently low specific energy $U=H/N$,
to a homogeneous phase (\(HP\)) at high energy.
For non zero values of \( A \), the two spatial directions are coupled.
Previous investigations were limited to the value \( A =1 
\)~\cite{anttor} and
a transition was also observed from a \( CP_1 \) to a \(HP\)
phase. This transition is first-order in the canonical ensemble, 
while microcanonical simulations are compatible with
a continuous transition associated with a negative specific heat regime. 
Both for $A=0$ and $A=1$, at low energies, 
all particles are trapped in a cluster, while, for sufficiently high energies, 
they are uniformly distributed in the cell.

In order to better investigate the origin of
ensemble inequivalence within a unique
framework, we have introduced
model (\ref{eqHA}) which allows, by continuously
varying parameter \( A\), to pass from
a situation where the microcanonical and canonical 
results coincide ($A=0$) to a situation where the two 
ensembles disagree over a finite energy range
near the transition.
As we will show, this model is
indeed richer than expected because it reveals
also more complicated transitions than those
previously studied in~\cite{anttor,antruf}.

Due to the long-range nature of the interactions,
the collective behaviour of the particles can be described in terms of 
the following mean-field vectors:
\( \overrightarrow{M}_{z}=\left( \left\langle \cos (z)\right\rangle _{N},
\left\langle \sin (z)\right\rangle _{N}\right) =
M_{z}\left( \cos (\phi _{z}),\sin (\phi _{z})\right)  \)
where  $\phi_z \in [0,\pi/2]$ and \( z=x \) 
or \( y \); \( \overrightarrow{P}_{z}=
\left( \left\langle \cos (z)\right\rangle _{N},
\left\langle \sin (z)\right\rangle _{N}\right) =
P_{z}\left( \cos (\psi _{z}),\sin (\psi _{z})\right)  \)
where $\psi_z \in [0,\pi/2]$ and \( z=x \pm y \). 
The average over all the particles
is indicated by \( \left\langle ...\right\rangle _{N} \).
It can be shown that on average 
\( M_{x} \approx M_{y} \approx M \) 
and \( P_{x+y} \approx P_{x-y} \approx P \) (for more
details see~\cite{anttor}).
Therefore, the potential energy can be rewritten, in the mean-field limit 
$N \to \infty$, as \( V_A=[2+A-2 M^{2}-A P^{2}]/2  \).

For \( U \approx 0 \) (or equivalently at low temperature), 
the system described by model (\ref{eqHA}) is in the 
\( CP_{1} \), particles have all the same location in a single point-like 
cluster and \( M\approx P \approx 1 \),
whereas at large enough energy (temperature) 
the system is in the \( HP \) and \( M\approx P=O(1/\sqrt{N}) \).
For sufficiently high values of \( A  > A_2 \sim 3.5 \),
a third intermediate phase \( CP_{2} \), exhibiting two clusters,  
appears. In this phase, due to the symmetric location of the two clusters
\( M \sim O(1/\sqrt{N}) \) and \( P \sim O(1) \).

\begin{figure}
\centerline{
{\psfig{figure=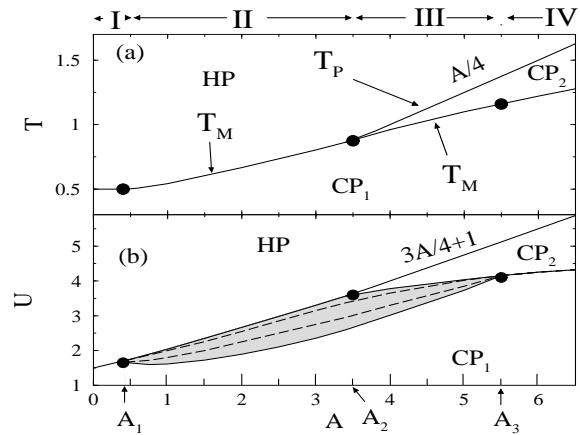,angle=270,width=7.5cm,height=6.0cm}}
}
\caption{\label{fig1} Phase diagrams of model (\ref{eqHA}) reporting
the transition
temperatures versus the coupling parameter $A$ (a) and the 
corresponding
specific energy 
$U = H/N$ versus  $A$ (b). The solid lines indicate the canonical 
transition 
lines and the dots the points where the nature of the transitions
change.
\protect\( A_{1}\protect \), \protect\( A_{2}\protect \) and 
\protect\( A_{3}\protect \)
are the threshold coupling constants that determine the transition 
scenario 
\protect\( I\to IV\protect \) displayed above the graphs. 
The grey shaded area
in (b) indicates the domain where two phases 
coexist.
Inside this area, the dashed curves are the two spinodal lines. }
\end{figure}

\vspace{0.3cm}

In the mean-field limit, the equilibrium properties of model 
(\ref{eqHA})
can be derived analytically within the canonical ensemble
following the approach of Ref.~\cite{anttor}. In particular,
the Helmholtz free energy reads as
\begin{equation}
\label{eqG}
\frac{F(M,P;T,A)}{T}=T(M^2+P^2)-\ln [T \,\, G(M,P;A)]
\end{equation}
\noindent 
with $G=\int _{0}^{2\pi }I_{0}
( M+\sqrt{2A} P \cos (s) ) \exp (M \cos(s)) ds$,
where \( I_{0} \) is the modified 
Bessel function of zero order. 
Since we are interested also in  metastable and unstable states
we will not restrict ourselves to the study of the lowest-lying minima 
of \( F \), but we will keep track of all the other extrema. 
Due to the $P \to -P$ symmetry of $F$, we can limit our analysis
to the $P \ge 0$ half-plane. The transition lines, obtained
by considering the absolute minima of $F$, are reported as solid 
lines in Fig.~\ref{fig1}. The minima can be easily associated
to the three observed phases, since $HP$ will correspond
to $M=P=0$, $CP_1$ to $|M|>0$,$|P|>0$ and $CP_2$ to $M=0$ and $|P|>0$.

Let us describe the observed phase transitions 
within the canonical ensemble with the help of Fig.~\ref{fig1}:
the line referred to as $T_{M}$ in the inset (a) indicates 
the transition temperatures where $M$ vanishes and
where the phase $CP_1$ looses its stability; while
the line $T_{P}$ is where $P \to 0$ and
the $CP_2$ leaves place to the $HP$.
The phase diagram $U$ versus $A$ reported in the inset (b) 
gives clearer hints for what concerns first order transitions,
indicating the corresponding energy jumps (latent heats).
Depending on the value of $A$, four different scenarios
can be distinguished.
\textbf{(I): $[ 0 \le A \le A_1 = 2/5]$ -}
in this case one observes a continuous transition
from a  $CP_1$ to a $HP$, the critical line is located
at $T_M=1/2$ ($U_M=3/2+A$). Along this line, at $A_1=2/5$,
there is a tricritical point : here the transition 
becomes discontinuous. 
\textbf{(II) : \([ A_{1}<A<A_{2}\approx 3.5 ]\) - } 
The transition between \( CP_{1} \) and \( HP \) is first order 
in all this range of parameters with a finite energy jump
(grey shaded area in Fig.~\ref{fig1} (b)).
\textbf{(III) : \([ A_{2}<A<A_{3}\approx 5.7 ]\) - } 
In this region the third phase begins to 
play a role and two successive transitions are observed:
first $CP_1$ disappears at $T_M$ via a first order transition
that gives rise to the biclustered phase $CP_2$, which
ends up in the $HP$ due to a continuous transition.
The critical line associated to this last
transition is $T_P =A/4$ ($U_P =3A/4+1$).
\textbf{(IV) : \([ A>A_{3} ]\) - } 
Finally, region (IV)  differs from
region (III) just for the nature of the
transition connecting the two clustered phases, that
becomes second order.

\begin{figure}
\centerline{
{\psfig{figure=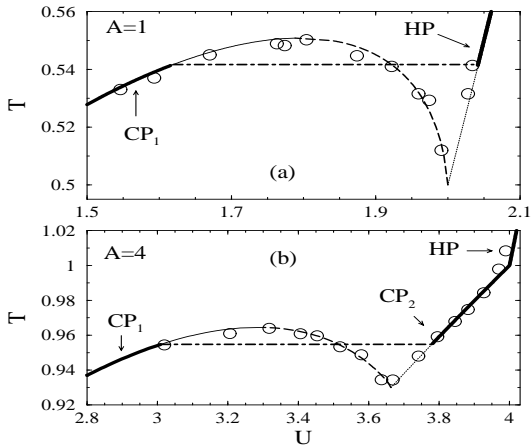,angle=270,width=7.0cm,height=6.0cm}}
}
\caption{\label{fig2} Canonical temperature-energy relation in the 
coexistence region
for A=1 (a) and A=4 (b). Lines indicate canonical analytical 
results, while
circles correspond to microcanonical simulations. Solid thick lines are
equilibrium results, solid thin lines metastable states
and dashed thin lines unstable states. The straight dashed thick
line is the Maxwell construction.
Figure (a) refers to a
first order transition from $CP_1$ to $HP$, (b) 
to discontinuous transition connecting the two clustered phases.
In (b) the second order transition from $CP_2$ to $HP$ is also shown.
The microcanonical results have been obtained via molecular
dynamics simulations of model (\ref{eqHA}) with $N=5,000$ 
particles~\protect\cite{md}.}
\end{figure}

Let us now concentrate on first order phase transitions, 
in particular we will compare microcanonical results
obtained via molecular dynamics simulations~\cite{md} with canonical
results. Outside the grey area in Fig.~\ref{fig1}(b) the microcanical
and canonical results coincide everywhere, but inside such
region strong discrepancies are observed. This can be clearly
seen in Fig. \ref{fig2}, where $T$  is plotted as a function
of $U$ near the discontinuous transition 
for two different values of $A$ (namely, $A=1$ and
$A=4$). Within the canonical ensemble, first order phase
transitions occur at a given temperature and
the coexistence region is bridged with a horizontal line 
($T=T_M$ in Fig. \ref{fig2}(a) and $T=T_P$ in \ref{fig2}(b)),
according to the  Maxwell construction.

However, this region is accessible via microcanonical
simulations and the corresponding equilibrium results
are shown as circles in Fig. \ref{fig2}.
Within this region these results disagree with the canonical ones, 
the transition is now continuous and a negative
specific heat region appears.
It is remarkable that
both in the transition connecting the clustered phase $CP_1$
to the $HP$ as well as in that connecting the two clustered phases
the essential features of the transition are the same.

\begin{figure}
\centerline{
{\psfig{figure=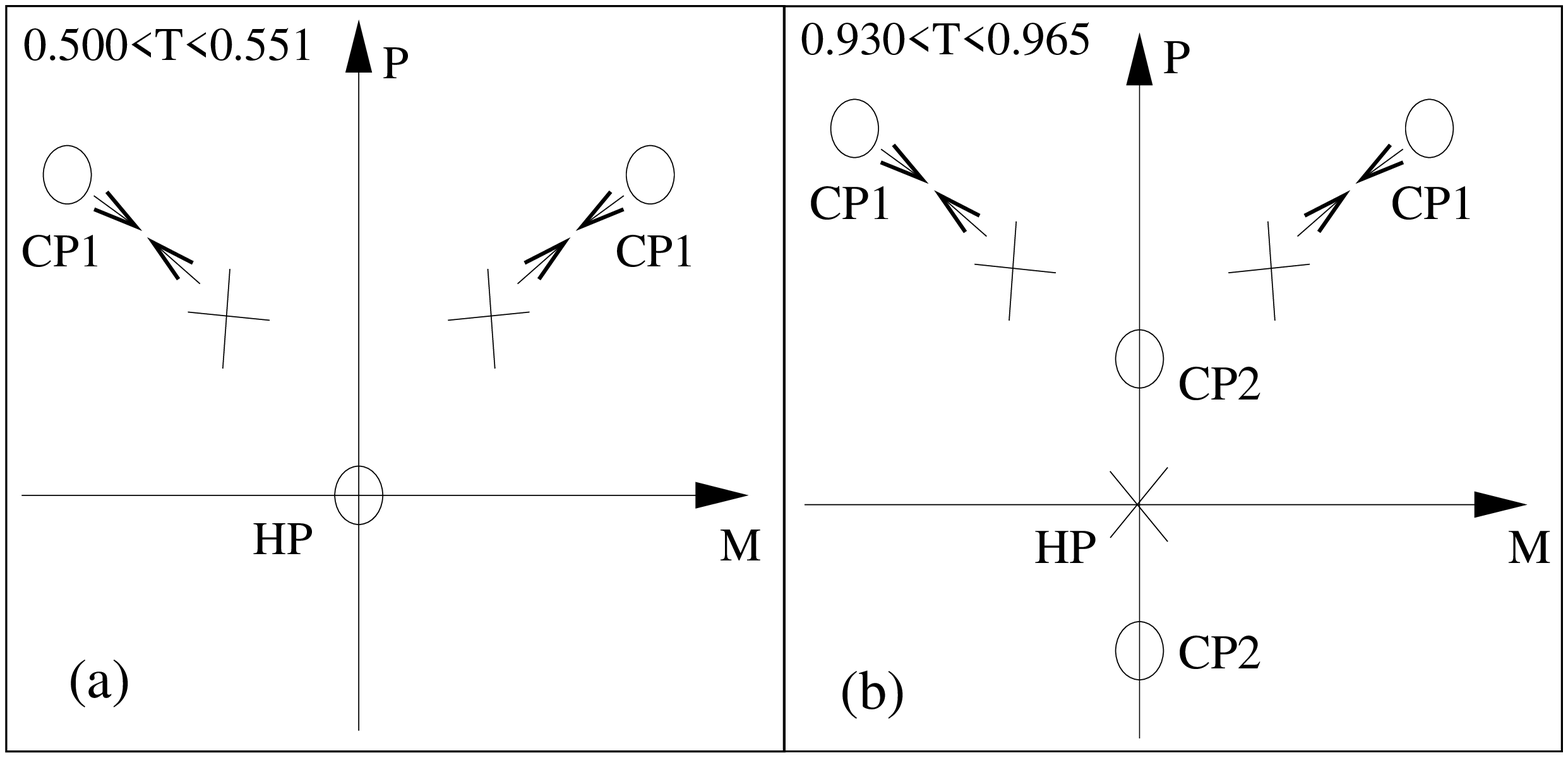,angle=0,width=8.0cm,height=3.5cm}}
}
\vskip 0.3 truecm
\caption{\label{fig3} Schematic representation
of the extrema of the free energy for a fixed temperature
in the intervals reported in the figures for $A=1$ (a) and $A=4$ (b). 
Circles denote the
positions of the minima, while the crosses represent
the saddles. The motion of the extrema for increasing 
temperatures is indicated by the arrows.
}
\end{figure}

We will now show that the microcanonical results
can be obtained even within the canonical ensemble
by considering also the relative minima and
the saddles of $F$. A schematic sketch of the relevant
extrema of $F$ in the $(M,P)$-plane close to
the two transitions discussed above
is reported in Fig.~\ref{fig3}.
Let us first consider the case $A=1$ for
$1/2 \leq T \leq 0.551$, that is reported in Fig.~\ref{fig3}(a).
For $T < 1/2$ the free energy has a unique minimum
corresponding to the $CP_1$ phase, while at $T=1/2$
a second relative minimum emerges at $M=P=0$
associated to a metastable
$HP$ state. Exactly at the same temperature 
also two symmetric saddles appear in the $(M,P)$-plane
inbetween the $CP_1$ and the $HP$ minima. For increasing
temperature the depth of the $CP_1$ minimum decreases
while that of the $HP$ minimum increases and, simultaneously,
the saddles and the $CP_1$ minima approach
each other. At $T=T_M \simeq 0.54$ the $CP_1$ and $HP$ mimima
reach exactly the same depth and this singles out the
canonical transition temperature. At $T = 0.551$ the
saddles and the $CP_1$ minima finally merge and disappear.
If the energy and temperature denoting these 
unstable and metastable states are reported in a graph
together with the equilibrium solutions, one observes that
they connect continuously the $CP_1$ equilibria to the $HP$ ones 
(see Fig.~\ref{fig2}). Moreover, as shown again in Fig.~\ref{fig2}(a),
the microcanonical results essentially coincide with the 
metastable and unstable canonical solutions in the coexistence region.
In particular, the negative specific heat regime
is associated to the saddles bridging
the metastable minima. The limits of existence
of these unstable states are tipically referred to as spinodals
and are reported in Fig.~\ref{fig1} as dashed lines
\cite{katz}.
A similar behaviour is observed for the $CP_1$ to $CP_2$
transition, in this case for $T < 0.93$ the unique minimum of
$F$ is again associated to the $CP_1$ and at $T=0.93$ two
symmetric minima appear on the $M=0$ axis with $|P| \ne 0$, 
and these metastable states are clearly associated to the $CP_2$.
Also two saddle points appear in the $F$ profile
separating the $CP_1$ minima from the metastable
$CP_2$ minimum in the positive $P$ semi-plane.
For $T>T_{P} \simeq 0.95$ the $CP_2$ minima become the
stable ones. For increasing temperatures the $CP_1$ minima
and the saddles approach each other and finally vanish
at the spinodal (located at $T \simeq 0.965$). For higher temperatures
the $CP_2$'s approach symmetrically the origin,
where a saddle (not involved in the first order transition)
is present corresponding to the $HP$ unstable phase.
At $T=T_{M} \simeq 1$ the two minima merge with the saddle at the
origin, giving rise to a unique minimum for $F$ corresponding to
a stable $HP$. This second transition is clearly continuous.
Also in this case, the microcanonical continuous transition is 
well reproduced by the metastable and unstable states 
involved in the discontinuous transition (see Fig.~\ref{fig2}(b)).

In conclusion, we have shown that a study of the absolute
and relative extrema of the Helmholtz free energy is sufficient 
to provide a complete microcanonical and canonical description of
the equilibrium behaviour of a $N$-body Hamiltonian 
with infinite-range attractive interaction.
Depending on the value of the tuning
parameter $A$ canonically first or second order transitions 
are observed. The comparison of the canonical results with
the microcanonical ones shows that ensemble inequivalence
occurs in proximity of canonically discontinuous transitions.
Irrespective of the phases involved in the transitions
a negative specific heat regime has been observed. In particular, 
this regime is always associated to the existence of a
spinodal region for the canonical solutions. For energy values
within this region microcanonically stable states corresponds to saddles
of the free energy. We have validated this scenario by considering
two different discontinuous transitions connecting the
one cluster phase either to the homogeneous one or to the
two cluster phase.

We believe that the results reported in this article are not
limited to systems with infinite-range interactions, but they
should be applicable also to systems with power-law decaying
potentials (as shown in \cite{tama}) and to
finite systems with short-ranged forces (see
\cite{gross_book}).
 
The MPIPKS Institute in Dresden is acknowledged
for providing computational facilities and F. Chauvet and H. Scherrer
for their advices. We thank J. Barr{\'e}, F. Bouchet, P.-H. Chavanis, 
T. Dauxois and D. Mukamel for useful discussions. 
This work is part of the MURST-COFIN00 research project
``Chaos and localization in classical and quantum mechanics''.

\end{multicols}


\begin{references}
\bibitem{md} All the simulation details can be found in \cite{anttor}.
\end{references}
\end{document}